\documentclass[journal]{IEEEtran}
\usepackage[cmex10]{amsmath}
\usepackage{amsthm}
\usepackage{amssymb}
\usepackage{graphicx}
\usepackage{cite}
\usepackage{color}
\usepackage{algorithmic}
\usepackage{algorithm}
\usepackage{bm} 
\usepackage{caption}

\begin{document}

\title{\LARGE{Optimization for Pinching Antennas System With Multiple Carriers and Rate Splitting Multiple Access
		
}}
\author{Peiyu Wang, Hong Wang,  Yaru Fu, and Rongfang Song \vspace{-0.8cm}
\thanks{
	This work was supported in part by National Natural Science
Foundation of China under Grant 62171235 and in part by Postgraduate Research \& Practice Innovation Program of Jiangsu
 Province under Grant KYCX25\_1096.   (\emph{Corresponding author: Rongfang Song}.)

Peiyu Wang, Hong Wang, and Rongfang Song are with the School of Communication and Information Engineering, Nanjing University of Posts and Telecommunications, Nanjing 210003, China (e-mail: wang\_py1999@163.com, wanghong@njupt.edu.cn, songrf@njupt.edu.cn).

Yaru Fu is with the School of Science and Technology, Hong Kong
Metropolitan University, Hong Kong SAR, China (e-mail: yfu@hkmu.edu.hk).}
}

\maketitle

\begin{abstract}
To meet the urgent demands for spectral efficiency and multi-user access in high-frequency application scenario for the sixth-generation wireless communication, this paper investigates a rate splitting multiple access (RSMA) system assisted by pinching antennas (PAs) with multiple waveguides and multiple carriers, aiming to maximize the overall system sum rate. To address the high sensitivity of high-frequency signals to PA movement in the overloaded scenarios, a two-stage PA position optimization method based on both path loss and phase shift error minimization is proposed under RSMA framework. Specifically, the first step is to perform coarse adjustment by minimizing large-scale path loss. Then, based on the derivation of a closed-form solution for the ideal phase shift in a single-user single-carrier case, the fine-grained positions of PAs are optimized via a one-dimensional line search to minimize the composite phase shift error across all users and carriers. In order to meet the quality of service requirements, the Lagrange dual method is employed to obtain the closed form of beamforming vectors after the PA positions are determined. Simulation results demonstrate that the proposed scheme achieves significant improvement in sum rate and confirm that RSMA exhibits stronger robustness to inaccurate PA positions caused by both discrete position channel estimation and physical hardware compared to other multiple-access techniques in PA-assisted systems. Furthermore, the results validate that fine-grained PA position adjustment is particularly crucial in high-frequency bands.
\end{abstract}


\section{Introduction}
The transition to the sixth generation mobile communication system (6G) necessitates the adoption of high-frequency bands to achieve greater bandwidth \cite{zhang20196g}. However, high-frequency transmission is inherently constrained by severe path loss, which markedly reduces the effective propagation range of signals. Consequently, the efficient utilization of high-frequency resources emerges as a pivotal research challenge in next-generation network design \cite{rangan2014millimeter}. 

Reconfigurable intelligent surface (RIS) is considered a promising technology for coverage extension; however, its multi-hop reflective architecture introduces significant cascaded path loss when redirecting signals to blind zones \cite{liu2021reconfigurable}. This characteristic makes RIS particularly unsuitable for high-frequency communication scenarios. On the other hand, movable antennas \cite{zhu2025tutorial} and fluid antennas \cite{new2024tutorial} enhance performance by expanding spatial degrees of freedom; however, the movement range of both antenna types is limited to several wavelengths, rendering them ineffective against large-scale fading.

In contrast to the aforementioned technologies, pinching antenna system (PASS) has attracted growing attention in recent research due to its capability to effectively reduce path loss and its flexibility in PA positioning. PASS is generally installed indoors and enables reconfiguration of the pinching antennas along the waveguide \cite{liu2025pinching}. This design enables PASS to significantly shorten the propagation distance and reduce path loss. Moreover, PASS can dynamically adjust PA positions to alter the phase shift of transmitted signals, thereby further enhancing signal strength. On the other hand, the PASS transmits high-frequency signals through waveguides and delivers them to near-field users via pinching antennas \cite{ding2025flexible}. Benefiting from the characteristics of waveguide transmission, the path loss in waveguide is minimal and often negligible \cite{yang2025pinching}. A study comparing the performance of PASS and RIS in high-frequency scenarios demonstrates that PASS outperforms RIS \cite{samy2025pinching}. 

The core problem in PASS lies in determining the positions of PAs to improve channel quality. Although both PASS and RIS can be regarded as forms of hybrid beamforming systems, they operate on fundamentally different principles. Unlike RIS, which directly control the phase of transmitted signals to reconfigure the channel, PASS reconfigures the channel by physically relocating its antenna elements, thereby indirectly influencing the phase distribution. Since each user exhibits distinct spatial geometry relative to different PA elements, the resulting variations in propagation distance and corresponding signal phase are non-uniform and user-dependent. This differs fundamentally from the globally uniform phase shift achieved by RIS. Consequently, RIS can apply a consistent and controllable phase adjustment for all users, while PA systems introduce a spatially varying and user-specific phase perturbation that depends strongly on the distribution of users. This necessitates a novel optimization scheme tailored for PA positioning. To reduce path loss, a user-position-averaged PA placement scheme is proposed in \cite{xie2025low}, which determines PA locations based on average user distributions. Unlike approaches that consider only path loss, a two-stage method introduced in \cite{xu2025rate} first determines the initial PA positions and then aligns their channel phases for fine-grained placement. Furthermore, the cross‑entropy optimization framework is applied in \cite{chen2025dynamic} to investigate both discrete and continuous PASS configurations respectively.

In the initial research phase, only a single-user scenario was considered for PASS \cite{xu2025rate}, without addressing multiple-access techniques. As the number of users increases, however, the limited number of available waveguides poses a challenge. Specifically, once the number of users exceeds that of waveguides, PASS can no longer reconfigure orthogonal channels, which significantly degrades the performance of space division multiple access (SDMA) systems.
 To address this overload scenario, recent research has shifted its focus primarily to non-orthogonal multiple access (NOMA)-assisted PASS. A central challenge in such systems lies in jointly addressing two critical issues: determining the decoding order and optimizing the positions of the PAs \cite{zhuo2025p}. Earlier studies focus on a distributed single-waveguide configuration, where decoding order is determined based on effective channel gains, and PA positioning is optimized via a matching-theory-based algorithm \cite{wang2025antenna}. Further extending this framework, multi-waveguide scenarios are investigated in \cite{wang2025antenna2}, in which the decoding order is jointly determined by interference, channel gains, and noise, while a novel game-theoretic algorithm is proposed to optimize PA placement. Unlike discrete PASS, a continuous PASS configurations have been proposed to further enhance system performance \cite{gan2025joint}. In this model, the decoding order within each cluster is directly determined by channel gains, and the PA position is optimized via a particle swarm optimization algorithm to minimize the transmit power. Although transmitting multiple signals via waveguide facilitates signal separation for successive interference cancellation (SIC) in power-domain NOMA, the bit error probability increases with growing number of users due to error propagation \cite{mao2022rate}. Moreover, in multiple-input multiple-output (MIMO) systems, Clerckx et al. \cite{clerckx2021noma} investigated system performance in overloaded scenarios. Simulations reveal that under overloaded conditions, NOMA may underperform SDMA due to its lack of spatial multiplexing gain. In contrast, rate splitting multiple access (RSMA) achieves superior performance by effectively balancing interference management and multiplexing gains.
 
Compared to SDMA and NOMA, RSMA serves as a bridge between the two \cite{mishra2022rate}, offering better compatibility with PASS. In overload scenarios where SDMA fails, RSMA efficiently utilizes channel state information by allocating private streams and managing interference through a common stream. In contrast to NOMA, RSMA does not require strict decoding order based solely on channel gains. Moreover, in multi-user settings, RSMA requires only a single layer of SIC, thereby reducing error propagation and hardware complexity \cite{mishra2021rate}.  A comparison between NOMA and RSMA in PASS is conducted \cite{hua2025content},\cite{tegos2025uplink}, demonstrating that RSMA outperforms both NOMA and SDMA. However, these studies only consider a single-waveguide configuration, which limits the efficient utilization of private streams. In contrast, \cite{wang2025sum} investigates an RSMA-assisted PASS system with multiple waveguides, where RSMA again exhibits superior performance over NOMA. 

 It must be emphasized that RSMA and PASS can leverage each other’s strengths to achieve mutual success. Specifically, in RSMA systems, the common rate is constrained by the user with the weakest channel gain.  The PASS can mitigate this limitation by adjusting the PA positions to shorten the effective transmission distance, thereby enhancing the channel conditions for that user. This demonstrates that PASS can effectively compensate for the inherent disadvantage of RSMA. On the other hand, although PA positions are continuously adjustable along the waveguide, only discrete channel state information (CSI) can be obtained in practice. This leads to inaccurate PA positioning, a problem that stems from the use of inaccurate CSI. Even with theoretically perfect CSI and the optimal PA positions, practical hardware limitations and deployment inaccuracies prevent its exact realization. Such deployment inaccuracies can also be regarded as a form of imperfect CSI, which similarly degrades system performance. Fortunately, compared to NOMA and SDMA, RSMA demonstrates greater robustness against imperfect CSI \cite{mao2022rate},\cite{clerckx2021noma}. Specifically, RSMA achieves enhanced interference management through common stream allocation, enabling it to effectively compensate for the inherent CSI imperfections in PASS.

It must be pointed out that PASS is typically deployed in high-frequency broadband systems, wherein the waveguide exhibits frequency-selective characteristics and the communication between each user and multiple pinching antennas involves  propagation delay differences, collectively resulting in a frequency-selective channel \cite{liu2025pinching2}. However, existing literature on PA placement optimization focuses only on single carrier scenarios and overlooks the characteristics of frequency selectivity.
In the multi-carrier scenario, more factors must be considered to optimize PA positioning. Specifically, even at the same position and distance, signals on different carriers experience distinct phase shifts due to their differing frequencies. These frequency-dependent phase variations further complicate the position optimization of the PAs. 

Based on the above discussion, an RSMA system assisted by multi-carrier PASS is proposed to maximize the sum rate. The main contributions of this paper are summarized as follows:

\begin{itemize}
	\item This paper presents the first study on the PA assisted RSMA system with multiple carriers. RSMA and PASS exhibit mutual enhancement: PASS improves the weak-user channel by shortening propagation distance, thereby elevating the RSMA common rate; conversely, RSMA compensates for PASS's inherent CSI inaccuracies through robust interference management via common stream allocation.


	\item Although PA systems share some similarities with RIS and hybrid beamforming systems, they cannot directly control phase shifts. Consequently, the optimization methods developed for RIS and hybrid beamforming systems are not directly applicable to PA systems. This necessitates the design of a novel optimization framework tailored for PA reconfiguration. Furthermore, in high-frequency scenarios, even a slight adjustment in the positions of the PA elements can significantly alter the signal phase, thereby substantially influencing overall system performance. To address this challenge, a two-step optimization scheme is proposed to determine PASS configuration. Specifically, the PAs’ initial positions are determined by minimizing the overall path loss. Then, for fine adjustment, unlike \cite{xu2025rate} which directly maximizes the sum rate, this work transforms the sum rate maximization objective into a phase shift optimization problem. In doing so, optimizing the phase shift indirectly improves the overall sum rate.
	\item This model considers a practical high-frequency broadband application scenario with multiple carriers. Unlike the single-carrier system, PAs introduce heterogeneous phase shifts across different subcarriers, making its optimal placement a critical problem. To handle this challenge , the fundamental relationship between the PAs' positions and the induced phase shifts is analyzed. For a given PA serving a single user on a specific subcarrier, a closed-form solution to the optimal channel phase shift is derived as a function of the PA's coordinates. Subsequently, a one-dimensional search scheme is proposed to determine the PA location that minimizes the aggregate phase shift error across all subcarriers for all users.
	\item The simulation results demonstrate that the proposed PA positioning scheme outperforms all baseline methods. Furthermore, they underscore the importance of fine-grained PA positioning even after the initial placement is determined. Additionally, the results confirm that the PASS framework effectively integrates with RSMA. Compared to other multiple access schemes,  RSMA in PASS delivers higher sum rate and greater robustness against imperfect CSI.
\end{itemize}

\begin{figure*}[h]
	\centering
	\includegraphics[width=\linewidth]{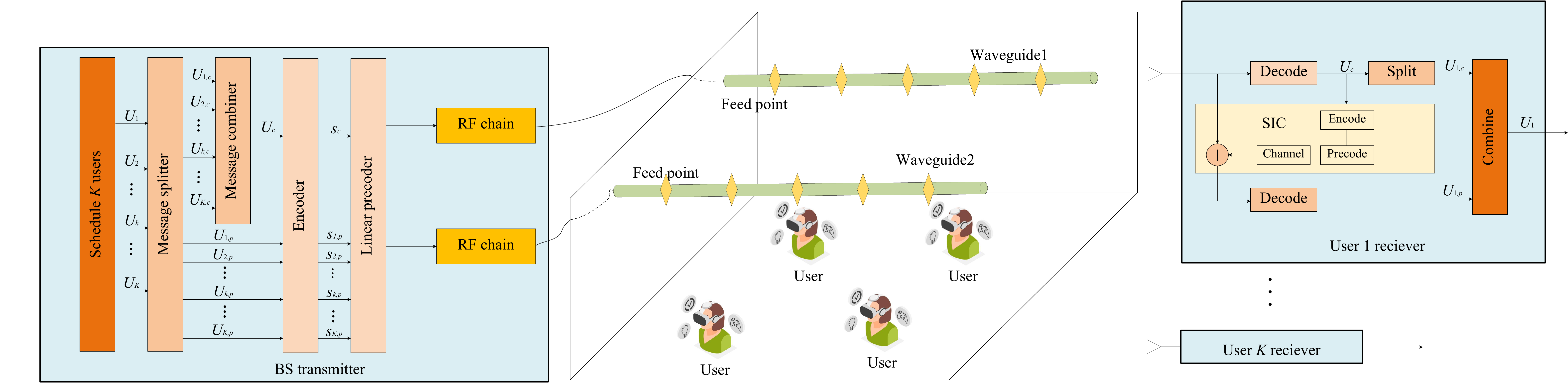}
	\caption{Illustration of downlink transmission for pinching antennas aided RSMA system.}\label{WANG_fig1}
\end{figure*}

\section{System Model}
This paper considers a downlink PA-assisted RSMA system operating over multiple carriers and employing multiple waveguides. The system comprises a base station (BS), $K$ users, $L$ carriers, and $M$ waveguides. Each waveguide is connected to a single radio frequency (RF) chain and equipped with $N$ mobile PAs. The PA positions can be adjusted according to the communication environment, subject to a minimum separation distance of $d_{\Delta} = \lambda/2$ between any two PAs. The $K$ users are randomly distributed within a square region of side length $D$, which is also the length of each waveguide.

\subsection{Channel Models}
 The channel model from the $n$-th PA on the $m$-th waveguide to the $k$-th user on the $l$-th carrier can be expressed as
\begin{align}\label{eq1}
{h}_{l,k,(m,n)}=\frac{\lambda_{l}}{4\pi}\cdot\frac{ e^{-j\frac{2\pi}{\lambda_{l}}|{\chi}^{PA}_{(m,n)}-{\chi}_{k}|}}{|{\chi}^{PA}_{(m,n)}-{\chi}_{k}|},
\end{align}
where $\lambda_{l}=\frac{c}{f_l}$ is the wavelength corresponding to the $l$-th carrier with $f_l$ and $c$ being the $l$-th carrier frequency and the speed of the light, respectively \cite{xu2025rate}, $|{\chi}^{PA}_{(m,n)}-{\chi}_{k}|=\sqrt{(x_{(m,n)}^{PA}-x_k)^2+(y_{(m,n)}^{PA}-y_k)^2+(z_{(m,n)}^{PA})^2}$ is the distance between the $n$-th PA on the $m$-th waveguide and the $k$-th user, $\chi_{(m,n)}^{PA}=(x_{(m,n)}^{PA},y_{(m,n)}^{PA},z_{(m,n)}^{PA})$, and $\chi_{k}=(x_{k},y_{k},0)$ is the corresponding Cartesian coordinates.

Let $g_{l,(m,n)}$ denote the propagation channel from the feed point to the $n$-th PA on the $m$-th waveguide for the $l$-th carrier, which can be expressed as

\begin{align}\label{eq1}
{g}_{l,(m,n)}=\sqrt{e^{-\alpha|\chi^{FP}_{m}-\chi^{PA}_{(m,n)}|}}e^{-j2\pi \frac{\eta_{eff}|\chi^{FP}_{m}-\chi^{PA}_{(m,n)}|}{\lambda_l}},
\end{align}
where $\sqrt{e^{-\alpha|\chi^{FP}_{m}-\chi^{PA}_{(m,n)}|}}$ is the in-waveguide propagation loss, $\alpha$ is the absorption coefficient, $\eta_{eff} > 1$ is the effective refractive index of the waveguide \cite{Wang2025Modeling}, $|{\chi}^{FP}_{m}-{\chi}^{PA}_{(m,n)}|=\sqrt{({x}^{FP}_{m}-{x}^{PA}_{(m,n)})^2+(y^{FP}_{m}-{y}^{PA}_{(m,n)})^2+({z}^{FP}_{m}-{z}^{PA}_{(m,n)})^2}$ is the distance between the feed point (FP) on the $m$-th waveguide and the $n$-th PA on the $m$-th waveguide, and $\chi_{m}^{FP}=(x_{m}^{FP},y_{m}^{FP},z_{m}^{FP})$ is the corresponding FP Cartersian coordinate.

 In this model, the non‑line‑of-sight (NLoS) link is neglected because the line‑of‑sight (LoS) signal is       sufficiently stronger than the NLoS component \cite{suzuki2022pinching}, \cite{rappaport201238}. This paper employs an equal power allocation strategy, distributing the transmit power of each waveguide equally among its corresponding PAs \cite{zeng2025energy}. The waveguide propagation loss is fully incorporated into the channel model via the channel coefficients $g_{l,(m,n)}$ \cite{wang2025antenna}. Thus, the equivalent channel of the $k$-th user can be given as
\begin{align}\label{channel}
\mathbf{h}_{l,k}=\tfrac{1}{\sqrt{N}}( \begin{smallmatrix}
\sum_{n=1}^{N}h_{l,k,(1,n)}g_{l,(1,n)} \\
\sum_{n=1}^{N}h_{l,k,(2,n)}g_{l,(2,n)}  \\
\vdots \\
\sum_{n=1}^{N}h_{l,k,(M,n)}g_{l,(M,n)}\\
\end{smallmatrix}),
\end{align}
where $\frac{1}{\sqrt{N}}$ represents the equal power allocation to each PA on a waveguide.

\subsection{Expression of Received Signal}

In RSMA systems, the data stream of each users consists of two parts, i.e., private stream and common stream. The private data stream  of each user can only be decoded by the corresponding user, while the private streams of other users are treated as noise. The common stream can be decoded by all users\cite{clerckx2023primer}. The received signal at user $k$ on carrier $l$ can be expressed as
\begin{align}\label{eq1}
{y}_{l,k}=&\underbrace{\mathbf{h}_{l,k}^H(\mathbf{w}_{l,k}^ps_{l,k}^p+\mathbf{w}_{l}^cs_l^c)}_{\text{desired signal}}+\underbrace{\sum_{i=1,i\neq k}^{K}\mathbf{h}_{l,k}^H\mathbf{w}_{l,i}^ps_{l,i}^p}_{\text{interference signal}}+\underbrace{{n}_{l,k}}_{\text{noise}},
\end{align}
where $\mathbf{w}_{l,k}^p$ is the private beamforming vector for user $k$ on carrier $l$, $\mathbf{w}_{l}^c$ is the common beamforming vector for all users on the same carrier $l$, both $s_{l,k}^p$ and $s_l^c$ are normalized power signals on carrier $l$, representing the private signal for user $k$ and the common signal, respectively, and ${n}_{l,k}$ is the additive white Gaussian noise with mean 0 and variance $\sigma^2$.

\subsection{Rate Expressions}

For downlink RSMA system, users first decode the common stream and treat all private streams as noise. Hence, the achievable rate for user $k$ on carrier $l$ decoding the common stream can be expressed as

\begin{align}
R_{l,k}^c=\log_2(1+\frac{|\mathbf{h}_{l,k}^{H}\mathbf{w}_{l}^c|^2}{\sum_{i=1}^{K}|\mathbf{h}_{l,k}^{H}\mathbf{w}_{l,i}^p|^2+\sigma^2_{l,k}}).
\end{align}

To ensure successful decoding of the common stream by all users, the actual common stream rate on the carrier $l$ must be bounded by the minimum achievable rate among all users, i.e., $R_l^c \leq min\{R_{l,1}^c,R_{l,2}^c,...,R_{l,K}^c\}$. Assume that the common rate allocated to user $k$ on carrier $l$ is denoted as $r_{l,k}^c$. It is worth noting that $R_{l,k}^c$ denotes the maximum achievable rate for the $k$-th user to decode the common stream, while $r_{l,k}^c$ represents the portion of the common rate allocated to that user. Consequently, the sum of all $r_{l,k}^c$ must satisfy $\sum_{k=1}^K r_{l,k}^c \le R_l^c$. 

After each user decodes the common stream, the SIC technology is used to remove the common stream from the received signal. Therefore, the achievable rate expression for the $k$-th user on carrier $l$ to decode the corresponding private stream can be expressed as

\begin{align}
R_{l,k}^p=\log_2(1+\frac{|\mathbf{h}_{l,k}^{H}\mathbf{w}_{l,k}^p|^2}{\sum_{i=1,i\neq k}^{K}|\mathbf{h}_{l,k}^{H}\mathbf{w}_{l,i}^p|^2+\sigma^2_{l,k}}).
\end{align}

The total achievable rate for the $k$-th user on the $l$-th carrier is the sum of its common rate and private rate, i.e., $R_{l,k} = r_{l,k}^c + R_{l,k}^p$.

\section{Problem and Solutions }

\subsection{Problem Formulation}
This paper addresses the sum rate maximization under a power budget constraint, with explicit considerations for quality of service (QoS) and rate allocation. The optimization problem is formulated as:
\begin{align}
\max_{\mathbf{w}_{l}^c,\mathbf{w}_{l,k}^p,r_{l,k}^c,x^{PA}_{(m,n)}}: &~ \sum_{l=1}^{L}R_l^c+\sum_{l=1}^{L}\sum_{k=1}^{K}R_{l,k}^p\nonumber\\
s.t. &~\sum_{l=1}^{L}R_{l,k}^p+\sum_{l=1}^{L}r_{l,k}^c \geq R_{min},\label{1req}\\
&~\sum_{k=1}^{K}r_{l,k}^c \leq R_{l}^c,\label{1com}\\
&~ R_{l}^c \leq min\{R_{l,1}^c,R_{l,2}^c,...,R_{l,K}^c\},\label{1comcons}\\
&~\sum_{l=1}^{L}||\mathbf{w}_l^c||^2+\sum_{l=1}^{L}\sum_{k=1}^{K}||\mathbf{w}_{l,k}^p||^{2}\leq P_{max},\label{1power}\\
&~|x^{PA}_{(m,n)}-x^{PA}_{(m,n-1)}| \geq d_{\Delta}, \label{nocoup}
\end{align}
where constraint (\ref{1req}) ensures the minimum transmission rate requirement $R_{min}$ is met, constraint (\ref{1com})  establishes an upper bound on the common rate allocation, constraint (\ref{1comcons}) ensures the common stream can be decoded by the all users, constraint (\ref{1power}) limits the total transmission power budget $P_{max}$, and constraint (\ref{nocoup}) ensures that the antenna elements remain uncoupled.

In order to maximize sum rate, the original problem is divided into two subproblems: PA positioning and waveguide beamfroming.
\subsection{PA Positioning Scheme}
In high‑frequency communication scenarios, the transmitted signal wavelength is on the order of millimeters, leading to high phase sensitivity to propagation distance that is challenging to PA positioning. To address this challenge, a two‑step PA positioning scheme is proposed, consisting of coarse adjustment followed by fine‑grained refinement. The coarse adjustment determines the initial placement by minimizing path loss and establishes a favorable basis for the subsequent fine‑grained refinement of PA positions. Once the initial positions are fixed, the path loss due to transmission distance can be treated as constant, thereby simplifying subsequent optimization. In the fine‑grained step, the optimal phase shift for a single‑user single‑carrier scenario is first derived in closed form. This result is then extended  to the general multi-user multi-carrier case, after which an optimization problem is formulated to minimize the composite phase shift error across all users and carriers, ultimately determining the final PA positions.

\subsubsection{Coarse Adjustment}
In order to reduce the path loss, the initial placement is designed to minimize the sum of distances between PAs and users. The coarse optimization problem can be expressed as
\begin{align}
\min_{ x_{(m,n)}^{PA} }:&\sum_{m=1}^M\sum_{n=1}^N\sum_{k=1}^K|\chi_{(m,n)}^{PA}-\chi_{k}|^2\nonumber\\
s.t.& ~0 \leq x_{(m,n)}^{PA}  \leq D\label{largesc}.
\end{align}

The coarse PA position optimization problem is convex and can thus be solved efficiently. After the center position of a PA is determined, the remaining PAs are placed to the left or right of this initial PA with a fixed inter‑element spacing of 
$D_\Delta$, subject to the minimum spacing constraint  $d_{\Delta}$.

\subsubsection{ Fine Adjustment}

After the initial PA positions are determined, the sum rate maximization objective function is reformulated to facilitate the fine‑grained adjustment of the PA positions. The private rate and common rate components are recast into a tractable auxiliary form, respectively. Then, the overall objective can be expressed as
\begin{align}
R=\sum_{k=1}^K\sum_{l=1}^L\max_{\alpha_{l,k}^p}\{&\text{log}_2(1+\alpha_{l,k}^p)-\tfrac{\alpha_{l,k}^p}{ln2}+\tfrac{(1+\alpha_{l,k}^p)\gamma_{l,k}^P}{ln2(1+\gamma_{l,k}^p)}\}\nonumber\\
+&\sum_{l=1}^L\min_{\alpha_{l,k}^c}\{\bar{R}_{l,1}^c,\bar{R}_{l,2}^c,...,\bar{R}_{l,K}^c\},\label{sum_rate_alpha}
\end{align}
where $\alpha_{l,k}$ is the auxiliary variable, $\gamma_{l,k}^p=\tfrac{|\mathbf{h}_{l,k}^H\mathbf{w}^p_{l,k}|^2}{\sum_{\widetilde{k}=1,\widetilde{k}\neq k}^K|\mathbf{h}_{l,k}^H\mathbf{w}^p_{l,\widetilde{k}}|^2}$, and $\bar{R}_{l,k}^c=\max\{\text{log}_2(1+\alpha_{l,k}^c)-\tfrac{\alpha_{l,k}^c}{ln2}+\tfrac{(1+\alpha_{l,k}^c)\gamma_{l,k}^c}{ln2(1+\gamma_{l,k}^c)}\}$ with $\gamma_{l,k}^c=\tfrac{|\mathbf{h}_{l,k}^H\mathbf{w}^c_{f}|^2}{\sum_{\widetilde{k}=1}^K|\mathbf{h}_{l,k}^H\mathbf{w}^p_{l,\widetilde{k}}|^2}$. The optimal $\widetilde{\alpha}_{l,k}$ can be obtained as ${\widetilde{\alpha}_{l,k}^p}=\gamma_{l,k}^p$ and ${\widetilde{\alpha}_{l,k}^c}=\gamma_{l,k}^c$.

\begin{proof}
	Taking $\alpha_{l,k}^p$ as an example, consider the expression   
	\begin{align}
 	\text{log}_2(1+\alpha_{l,k}^p) - \tfrac{\alpha_{l,k}^p}{\ln 2} + \tfrac{(1+\alpha_{l,k}^p)\gamma_{l,k}^p}{\ln 2(1+\gamma_{l,k}^p)}
	\end{align}
	 denoted as $\mathfrak{R}_{l,k}^p$. The partial derivative of  $\mathfrak{R}_{l,k}^p$ with respect to $\alpha_{l,k}^p$ is 
	 
	\begin{align}
	\frac{\partial\mathfrak{R}_{l,k}^p}{\partial\alpha_{l,k}^p}=\frac{1}{\ln2(1+\alpha_{l,k}^p)}-\frac{1}{\ln2}+\frac{\gamma_{l,k}^p}{\ln2(1+\gamma_{l,k}^p)}. 
	\end{align}
	Set $\frac{\partial \mathfrak{R}_{l,k}^p}{\partial \alpha_{l,k}^p}=0$
	to maximize $\mathfrak{R}_{l,k}^p$ ,which yields the optimal auxiliary variable $\widetilde{\alpha}_{l,k}^p=\gamma_{l,k}^p$. Following the same approach, ${\widetilde{\alpha}_{l,k}^c} = \gamma_{l,k}^c$ can be obtained.
\end{proof}

In RSMA systems, the common rate is constrained by the user with the weakest channel. Furthermore, the mathematical form of the common rate expression complicates direct optimization. To address this, a weight of  $\frac{1}{K}$ is introduced for the common rate in the formulation, which appropriately captures its contribution to the overall sum rate. The rationale for this weighting stems from the fact that common data can more effectively manage inter‑user interference than private data. Therefore, to maximize the total sum rate, the common rate should be made as large as possible. However, because the common rate is bounded by the worst‑channel user, improving the common rate for every user effectively elevates the overall common‑rate limit. Moreover, the weighted common rate serves as an upper bound to the original common‑rate expression, ensuring that the optimization remains feasible and well‑posed. Substituting $\widetilde{\alpha}_{l,k}^p$ and $\widetilde{\alpha}_{l,k}^c$ into (\ref{sum_rate_alpha}) and noting that both 
$\widetilde{\alpha}_{l,k}^p$ and $\widetilde{\alpha}_{l,k}^c$ are now fixed, the first term 
$\text{log}_2(1+\widetilde{\alpha}_{l,k}^p)$ and the second term $\tfrac{\widetilde{\alpha}_{l,k}^p}{\ln 2}$ become constants. Remove these constant terms, and the sum rate expression can be simplified to
  \begin{align}
R=\sum_{l=1}^L\sum_{k=1}^K\widetilde{R}_{l,k}^p+\sum_{l=1}^L\sum_{k=1}^K\tfrac{1}{K}\widetilde{R}_{l,k}^c,
\end{align}
where
$\widetilde{R}_{l,k}^p=\tfrac{(1+\widetilde{\alpha}_{l,k}^p)|\mathbf{h}_{l,k}^H\mathbf{w}_{l,k}^p|^2}{\sum_{\widetilde{k}=1}^K|\mathbf{h}_{l,k}^{H}\mathbf{w}^p_{l,\widetilde{k}}|^2+\sigma^2}$ and $\widetilde{R}_{l,k}^c=\tfrac{(1+\widetilde{\alpha}_{l,k}^c)|\mathbf{h}_{l,k}^H\mathbf{w}_{l}^c|^2}{|\mathbf{h}_{l,k}^{H}\mathbf{w}^c_{l}|^2+\sum_{\widetilde{k}=1}^K|\mathbf{h}_{l,k}^{H}\mathbf{w}^p_{l,\widetilde{k}}|^2+\sigma^2}$.

Since the channel depends on the positions of the PAs, both the numerator and denominator of the rate expression involve channel-related terms, making the problem difficult to solve directly. To simplify the formulation, fractional programming is applied to transform $\widetilde{R}_{l,k}^p$ and $\widetilde{R}_{l,k}^c$ into the following more tractable forms \cite{shen2018fractional}. The sum rate expression can be transformed into
\begin{align}
\widetilde{R}_{l,k}^p=&\text{Re}\{\sqrt{1+\widetilde{\alpha}_{l,k}^p}((\xi_{l,k}^p)^*\mathbf{h}_{l,k}^{H}\mathbf{w}_{l,k}^p+(\mathbf{w}_{l,k}^p)^H\mathbf{h}_{l,k}\xi_{l,k}^p)\label{r_p_au}\nonumber\\
&-|\xi_{l,k}^p|^2(\sum_{\widetilde{k}=1}^K|\mathbf{h}_{l,k}^{H}\mathbf{w}_{l,\widetilde{k}}^p|^2+\sigma^2)\},\\
\widetilde{R}_{l,k}^c=&\text{Re}\{\sqrt{1+\widetilde{\alpha}_{l,k}^c}((\xi_{l,k}^c)^*\mathbf{h}_{l,k}^{H}\mathbf{w}_{l}^c+(\mathbf{w}_{l}^c)^H\mathbf{h}_{l,k}\xi_{l,k}^c)\nonumber\\
&-|\xi_{l,k}^c|^2(|\mathbf{h}_{l,k}^{H}\mathbf{w}^c_{l}|^2+\sum_{\widetilde{k}=1}^K|\mathbf{h}_{l,k}^{H}\mathbf{w}_{l,\widetilde{k}}^p|^2+\sigma^2)\},\label{r_c_au}
\end{align}
where the $\xi_{l,k}^p$ and $\xi_{l,k}^c$ are auxiliary variables. 

The optimal auxiliary variables are obtained by differentiating (\ref{r_p_au}) and (\ref{r_c_au}) with respect to $\xi_{l,k}^p$ and $\xi_{l,k}^c$, respectively, which gives
 \begin{align}
\frac{\partial\widetilde{R}_{l,k}^p}{\partial\xi_{l,k}^p}=&2\sqrt{1+\widetilde{\alpha}_{l,k}^p}\mathbf{h}_{l,k}^{H}\mathbf{w}_{l,k}^p
-2\xi_{l,k}^p(\sum_{\widetilde{k}=1}^K|\mathbf{h}_{l,k}^{H}\mathbf{w}_{l,\widetilde{k}}^p|^2+\sigma^2),\label{xi_p}\\
\frac{\partial\widetilde{R}_{l,k}^c}{\partial\xi_{l,k}^c}=&2\sqrt{1+\widetilde{\alpha}_{l,k}^c}(\mathbf{h}_{l,k}^{H}\mathbf{w}_{l}^c+(\mathbf{w}_{l}^c)^H)\label{xi_c}\nonumber\\
&-2\xi_{l,k}^c(|\mathbf{h}_{l,k}^{H}\mathbf{w}^c_{l}|^2+\sum_{\widetilde{k}=1}^K|\mathbf{h}_{l,k}^{H}\mathbf{w}_{l,\widetilde{k}}^p|^2+\sigma^2).\
 \end{align}
Setting the derivatives in (\ref{xi_p}) and (\ref{xi_c}) to zero yields the optimal auxiliary variables, denoted as 
  $\widetilde{\xi}_{f,k}^p=\tfrac{\sqrt{1+\widetilde{\alpha}_{f,k}^p}\mathbf{h}_{f,k}^H\mathbf{w}_{f,k}^p}{\sum_{\widetilde{k}=1}^K|\mathbf{h}_{f,k}^{H}\mathbf{w}_{f,\widetilde{k}}^p|^2+\sigma^2}$ and $\widetilde{\xi}_{f,k}^c=\tfrac{\sqrt{1+\widetilde{\alpha}_{f,k}^c}\mathbf{h}_{f,k}^H\mathbf{w}_{f}^c}{|\mathbf{h}_{f,k}^{H}\mathbf{w}^c_{f}|^2+\sum_{\widetilde{k}=1}^K|\mathbf{h}_{f,k}^{H}\mathbf{w}_{f,\widetilde{k}}^p|^2+\sigma^2}$.
  
After applying fractional programming to the sum‑rate expression and determining the auxiliary variables, the only remaining optimization variables are the positions of the PAs. To perform fine‑grained position adjustment, the PAs are optimized sequentially. Specifically, with the positions of all other PAs held fixed, the position of the $n$-th PA on the $m$-th waveguide for carrier $l$ is optimized. During this single PA optimization, the channel expression can be separated into a term that depends on the position variable and a term that is independent of it; the latter can be treated as constant. This is done by decomposing the total channel as $\mathbf{h}_{l,k} = \mathbf{h}_{l,k,(m,n)} + \widetilde{\mathbf{h}}_{l,k,(m,n)}$, where $\mathbf{h}_{l,k,(m,n)} = [0,\dots,g_{f,k,(m,n)},\dots, 0]^T$ denotes the channel from the $n$-th PA on the $m$-th waveguide to user $k$ with carrier $l$, and $\widetilde{\mathbf{h}}_{l,k,(m,n)}=[\sum_{i=1}^Ng_{l,k,(1,i)},\dots,\sum_{i=1,i\neq n}^Ng_{l,k,(m,i)},\dots, \sum_{i=1}^Ng_{l,k,(M,i)}]^T$ represents the remaining channel after removing $\mathbf{h}_{l,k,(m,n)}$. Then, the sum rate expression can be further written as:

 \begin{align}
	R=\sum_{l=1}^L\sum_{k=1}^K\widetilde{R}_{l,k,(m,n)}^p+\sum_{l=1}^L\sum_{k=1}^K\tfrac{1}{K}\widetilde{R}_{l,k,(m,n)}^c,
\end{align}
where $\widetilde{R}_{l,k,(m,n)}^p$ and $\widetilde{R}_{l,k,(m,n)}^c$ represent the rates after separating out the channel component $\mathbf{h}_{l,k,(m,n)}$, and are given by
\begin{align}
	&\widetilde{R}_{l,k,(m,n)}^p=\label{r_p_frac}\\
	&\text{Re}\{\sqrt{1+\widetilde{\alpha}_{l,k}^p}((\widetilde{\xi}_{l,k}^p)^*(\mathbf{h}_{l,k,(m,n)}^{H}\mathbf{w}_{l,k}^p+\mathbf{\widetilde{h}}_{l,k,(m,n)}^{H}\mathbf{w}_{l,k}^p)\nonumber\\
	&+((\mathbf{w}_{l,k}^p)^H\mathbf{h}_{l,k,(m,n)}+(\mathbf{w}_{l,k}^p)^H\widetilde{\mathbf{h}}_{l,k,(m,n)})\widetilde{\xi}_{l,k}^p)\nonumber\\
	&-|\widetilde{\xi}_{l,k}^p|^2(\sum_{\widetilde{k}=1}^K|\mathbf{h}_{l,k,(m,n)}^{H}\mathbf{w}_{l,\widetilde{k}}^p+\mathbf{\widetilde{h}}_{l,k,(m,n)}^{H}\mathbf{w}_{l,\widetilde{k}}^p|^2+\sigma^2)\},\nonumber\\
	&\widetilde{R}_{l,k,(m,n)}^c=\label{r_c_frac}\\
	&\text{Re}\{\sqrt{1+\widetilde{\alpha}_{l,k}^c}((\widetilde{\xi}_{l,k}^c)^*(\mathbf{h}_{l,k,(m,n)}^{H}\mathbf{w}_{l}^c+\mathbf{\widetilde{h}}_{l,k,(m,n)}^{H}\mathbf{w}_{l}^c)\nonumber\\
	&+((\mathbf{w}_{l}^c)^H\mathbf{h}_{l,k,(m,n)}+(\mathbf{w}_{l}^c)^H\mathbf{\widetilde{h}}_{l,k,(m,n)})\widetilde{\xi}_{l,k}^c)\nonumber\\
	&-|\widetilde{\xi}_{l,k}^c|^2(|\mathbf{h}_{l,k,(m,n)}^{H}\mathbf{w}^c_{l}+\mathbf{\widetilde{h}}_{l,k,(m,n)}^{H}\mathbf{w}^c_{l}|^2+\nonumber\\
	&\sum_{\widetilde{k}=1}^K|\mathbf{h}_{l,k,(m,n)}^{H}\mathbf{w}_{l,\widetilde{k}}^p+\mathbf{\widetilde{h}}_{l,k,(m,n)}^{H}\mathbf{w}_{l,\widetilde{k}}^p|^2+\sigma^2)\}\nonumber.
\end{align}

 $\widetilde{R}_{l,k,(m,n)}^p$ and $\widetilde{R}_{l,k,(m,n)}^c$ can be further simplified. For example, in (\ref{r_p_frac}), the first term of $\widetilde{R}_{l,k,(m,n)}^p$ can be simplified as
\begin{align}
	&2\sqrt{1+\widetilde{\alpha}_{f,k}^p}\text{Re}\{\mathbf{h}_{l,k,(m,n)}^{H}\mathbf{w}_{l,k}^p(\xi_{l,k}^p)^*+\mathbf{\widetilde{h}}_{l,k,(m,n)}^{H}\mathbf{w}_{l,k}^p(\widetilde{\xi}_{l,k}^p)^*\}.
\end{align}
The squared term in the second component of $\widetilde{R}_{l,k,(m,n)}^p$ can be expanded as follows
\begin{align}
	&|\widetilde{\xi}_{l,k}^p|^2(\mathbf{h}_{l,k,(m,n)}^{H}\mathbf{W}_{l}^p\mathbf{h}_{l,k,(m,n)}+2\text{Re}\{\mathbf{h}_{l,k,(m,n)}^{H}\mathbf{W}_{l}^p\mathbf{\widetilde{h}}_{l,k,(m,n)}\}\nonumber\\
	&+\mathbf{\widetilde{h}}_{l,k,(m,n)}\mathbf{W}_{l}^p\mathbf{\widetilde{h}}_{l,k,(m,n)})+|\widetilde{\xi}_{l,k}^p|^2\sigma^2\nonumber,
\end{align}
where $\mathbf{W}_{l}^p=\sum_{\widetilde{k}=1}^K\mathbf{w}_{l,\widetilde{k}}^p(\mathbf{w}_{l,\widetilde{k}}^p)^H$.

After simplifying the first and second terms individually, the complete expression for $\widetilde{R}_{l,k,(m,n)}^p$ can be rewritten as
\begin{align}
\widetilde{R}_{l,k,(m,n)}^p=&2\text{Re}\{\mathbf{h}_{l,k,(m,n)}^H(\sqrt{1+\widetilde{\alpha}_{l,k}^p}\mathbf{w}_{l,k}^p(\widetilde{\xi}_{l,k}^p)^*\nonumber\\
&+|\widetilde{\xi}_{l,k}^p|^2\mathbf{W}_{l}^p\mathbf{\widetilde{h}}_{l,k,(m,n)})\}\nonumber\\
&+|\widetilde{\xi}_{l,k}^p|^2\mathbf{h}_{l,k,(m,n)}^H\mathbf{W}_{l}^p\mathbf{h}_{l,k,(m,n)}+\widetilde{c}_{l,k,(m,n)}^p\label{rp_frac},
\end{align}
where  $\widetilde{c}_{l,k,(m,n)}^p=2\text{Re}\{\sqrt{1+\widetilde{\alpha}_{l,k}^p}\mathbf{\widetilde{h}}_{l,k,(m,n)}^{H}\mathbf{w}_{l,k}^p(\widetilde{\xi}_{l,k}^p)^*\}+|\widetilde{\xi}_{l,k}^p|^2\mathbf{\widetilde{h}}_{l,k,(m,n)}\mathbf{W}_{l}^p\mathbf{\widetilde{h}}_{l,k,(m,n)}$ can be treated as a constant.

Following a similar approach, the common‑rate expression simplifies to
\begin{align}
	\widetilde{R}_{l,k,(m,n)}^c=&2\text{Re}\{\mathbf{h}_{l,k,(m,n)}^H(\sqrt{1+\widetilde{\alpha}_{l,k}^c}\mathbf{w}_{l}^c(\widetilde{\xi}_{l,k}^c)^*\nonumber\\
	&+|\widetilde{\xi}_{l,k}^c|^2\mathbf{W}_{l}^c\mathbf{\widetilde{h}}_{l,k,(m,n)})\}\nonumber\\
	&+|\widetilde{\xi}_{l,k}^c|^2\mathbf{h}_{l,k,(m,n)}^H\mathbf{W}_{l}^c\mathbf{h}_{l,k,(m,n)}+\widetilde{c}_{l,k,(m,n)}^c\label{rc_frac},
\end{align}
where $\widetilde{c}_{l,k,(m,n)}^c=2\text{Re}\{\sqrt{1+\widetilde{\alpha}_{l,k}^c}\mathbf{\widetilde{h}}_{l,k,(m,n)}^{H}\mathbf{w}_{l,k}^c(\widetilde{\xi}_{l,k}^c)^*\}+|\widetilde{\xi}_{l,k}^c|^2\mathbf{\widetilde{h}}_{l,k,(m,n)}\mathbf{W}_{l}^c\mathbf{\widetilde{h}}_{l,k,(m,n)}$ can be treated as a constant with $\mathbf{W}_{l}^c=\sum_{\widetilde{k}=1}^K\mathbf{w}_{l,\widetilde{k}}^p(\mathbf{w}_{l,\widetilde{k}}^p)^H+\mathbf{w}_{l}^c(\mathbf{w}_{l}^c)^H$.

After the private and common rate expressions are simplified, the channel can be further decomposed into the path loss and phase shift components. Although both components are influenced by the PA positions, the phase shift component is affected more significantly than the path loss. Specifically, in high‑frequency scenarios, a small displacement of a PA can induce a substantial phase shift, whereas the propagation distance—and consequently the path loss—can be regarded as approximately constant. This approximation is valid because large‑scale fading is predetermined and the distance‑induced path loss remains nearly invariant over a fine‑grained time interval. Therefore, the channel can be further decomposed into  distance and  phase shift components, i.e., $\mathbf{h}_{l,k,(m,n)}=\text{diag}(\mathbf{d}_{l,k,(m,n)})\bm{\varphi}_{l,k,(m,n)}$, where 
\begin{align}
	&\mathbf{d}_{l,k,(m,n)}=[0,\dots,\tfrac{\lambda_l\sqrt{e^{-\alpha|\chi^{FP}_{m}-\chi^{PA}_{n,m}|}}}{4\pi|{\chi}^{PA}_{n,m}-{\chi}_{k}|},\dots,0]^T,\
\end{align}
and 
\begin{align}
	&\bm{\varphi}_{l,k,(m,n)}=\nonumber\\
	&~~~[0,\dots,e^{-j\frac{2\pi}{\lambda_l}(|{\chi}^{PA}_{n,m}-{\chi}_{k}|+\eta_{eff}|\chi^{FP}_{m}-\chi^{PA}_{(m,n)}}|),\dots,0]^T,\
\end{align}
respectively denote the distance and phase shift components between the $n$-th PA on the $m$-th waveguide and the $k$-th user on the $l$-th carrier.

 Then, the first term of the private rate expression in (\ref{rp_frac}) can be rewritten as
\begin{align}
&\text{Re}\{\mathbf{h}_{l,k,(m,n)}^H(\sqrt{1+\widetilde{\alpha}_{l,k}^p}\mathbf{w}_{l,k}^p(\widetilde{\xi}_{l,k}^p)^*+|\widetilde{\xi}_{l,k}^p|^2\mathbf{W}_{l}^p\mathbf{\widetilde{h}}_{l,k,(m,n)})\}\nonumber\\
=&\text{Re}\{\bm{\varphi}_{l,k,(m,n)}^H\text{diag}(\mathbf{d}_{l,k,(m,n)})^H(\sqrt{1+\widetilde{\alpha}_{l,k}^p}\mathbf{w}_{l,k}^p(\widetilde{\xi}_{l,k}^p)^*\nonumber\\
&+|\widetilde{\xi}_{l,k}^p|^2\mathbf{W}_{l}^p\mathbf{\widetilde{h}}_{l,k,(m,n)})\}\nonumber\\
=&\text{Re}\{\bm{\varphi}_{l,k,(m,n)}^H\mathbf{q}_{l,k,(m,n)}^p\}\nonumber\\
=&\text{Re}\{(\varphi_{l,k,(m,n)}^{(m,1)})^*q_{l,k,(m,n)}^{p,(m,1)}+\sum_{j=1,j \neq m}^M(\varphi_{l,k,(m,n)}^{(j,1)})^*q_{l,k,(m,n)}^{p,(j,1)}\},
\end{align}
where $\mathbf{q}_{l,k,(m,n)}^p=\text{diag}(\mathbf{d}_{l,k,(m,n)})^H(\sqrt{1+\widetilde{\alpha}_{l,k}^p}\mathbf{w}_{l,k}^p(\widetilde{\xi}_{l,k}^p)^*+|\widetilde{\xi}_{l,k}^p|^2\mathbf{W}_{l}^p\mathbf{\widetilde{h}}_{l,k,(m,n)})$, and $\varphi_{l,k,(m,n)}^{(m,1)}$ and $q_{l,k,(m,n)}^{p,(m,1)}$ are the $m$-th elements of the vectors $\bm{\varphi}_{l,k,(m,n)}$ and $\mathbf{q}_{l,k,(m,n)}^{p}$, respectively.

After removing the irrelevant terms, the second term of the private rate expression in (\ref{rp_frac}) simplifies to
\begin{align}
&|\widetilde{\xi}_{l,k}^p|^2\mathbf{h}_{l,k,(m,n)}^H\mathbf{W}_{l}^p\mathbf{h}_{l,k,(m,n)}\nonumber\\
=&|\widetilde{\xi}_{l,k}^p|^2\bm{\varphi}_{l,k,(m,n)}^H\text{diag}(\mathbf{d}_{l,k,(m,n)})^H\mathbf{W}_{l}^p\nonumber\\
&\times\text{diag}(\mathbf{d}_{l,k,(m,n)})\bm{\varphi}_{l,k,(m,n)}\nonumber\\
=&|\widetilde{\xi}_{l,k}^p|^2\bm{\varphi}_{l,k,(m,n)}^H\mathbf{\widetilde{W}}_{l}^p\bm{\varphi}_{l,k,(m,n)}\label{p_second},
\end{align}
where $\mathbf{\widetilde{W}}_{l}^p=\text{diag}(\mathbf{d}_{l,k,(m,n)})^H\mathbf{W}_{l}^p\text{diag}(\mathbf{d}_{l,k,(m,n)})$.

Following the simplification above, (\ref{p_second}) can be further expressed as
\begin{align}
&\bm{\varphi}_{l,k,(m,n)}^H\mathbf{\widetilde{W}}_{l}^p\bm{\varphi}_{l,k,(m,n)}\label{phaseshift}\\
&=\sum_{i=1}^M\sum_{j=1}^M(\varphi_{l,k,(m,n)}^{(i,1)})^*\widetilde{w}_{l}^{p,(i,j)}\varphi_{l,k,(m,n)}^{(j,1)}\nonumber\\
&=(\varphi_{l,k,(m,n)}^{(m,1)})^*\widetilde{w}_{l}^{p,(m,m)}\varphi_{l,k,(m,n)}^{(m,1)}\nonumber\\
~&+\sum_{i=1,i\neq m}^M\sum_{j=1}^M(\varphi_{l,k,(m,n)}^{(i,1)})^*\widetilde{w}_{l}^{p,(i,j)}\varphi_{l,k,(m,n)}^{(j,1)}\nonumber\\
~&+\sum_{i=1}^M\sum_{j=1,j\neq m}^M(\varphi_{l,k,(m,n)}^{(i,1)})^*\widetilde{w}_{l}^{p,(i,j)}\varphi_{l,k,(m,n)}^{(j,1)}\nonumber\\
~&+\sum_{i=1,\neq m}^M\sum_{j=1,\neq m}^M(\varphi_{l,k,(m,n)}^{(i,1)})^*\widetilde{w}_{l}^{p,(i,j)}\varphi_{l,k,(m,n)}^{(j,1)},\
\end{align}
where $\widetilde{w}_{l}^{p,(i,j)}$ is the $i$-th row and the $j$-th column element of $\mathbf{\widetilde{W}}_{l}^p$. 

Given that only the $m$-th element of $\bm{\varphi}_{l,k,(m,n)}^H$ is non-zero, (\ref{phaseshift}) can be simplified to
\begin{align}
&\bm{\varphi}_{l,k,(m,n)}^H\mathbf{\widetilde{W}}_{l}^p\bm{\varphi}_{l,k,(m,n)}\nonumber\\
&=(\varphi_{l,k,(m,n)}^{(m,1)})^*\widetilde{w}_{l}^{p,(m,m)}\varphi_{l,k,(m,n)}^{(m,1)}=\widetilde{w}_{l}^{p,(m,m)}.\
\end{align}
The equality $(\varphi_{l,k,(m,n)}^{(m,1)})^*\widetilde{w}_{l}^{p,(m,m)}\varphi_{l,k,(m,n)}^{(m,1)}=\widetilde{w}_{l}^{p,(m,m)}$ holds because $\varphi_{l,k,(m,n)}^{(m,1)}$ is a complex number with unit modulus, i.e., $(\varphi_{l,k,(m,n)}^{(m,1)})^*\varphi_{l,k,(m,n)}^{(m,1)}=1$.

Based on the simplifications of the first and second terms, the problem of maximizing the private rate of user $k$ on the carrier $l$ with respect to the phase shift induced by the $n$-th PA on the $m$-th waveguide can be equivalently transformed into 
\begin{align}
\max_{\varphi}&~{\widetilde{R}_{l,k,(m,n)}^p}=\nonumber\\
\max_{\varphi}&~2\text{Re}\{(\varphi_{l,k,(m,n)}^{(m,1)})^*q_{l,k,(m,n)}^{p,(m,1)}\nonumber\\
&+\sum_{j=1,j \neq m}^M(\varphi_{l,k,(m,n)}^{(j,1)})^*q_{l,k,(m,n)}^{p,(j,1)}\}\nonumber\\
&\times|\widetilde{\xi}_{l,k}^p|^2\widetilde{w}_{l}^{p,(m,m)}+\widetilde{c}_{l,k,(m,n)}^p
\end{align}
After removing constant terms $\widetilde{c}_{l,k,(m,n)}^p$, the final form of the private rate with respect to the PA position induced phase shift can be expressed as
\begin{align}
\max_{\varphi}~~ 2\text{Re}\{(\varphi_{l,k,(m,n)}^{(m,1)})^*q_{l,k,(m,n)}^{p,(m,1)}\}|\widetilde{\xi}_{l,k}^p|^2\widetilde{w}_{l}^{p,(m,m)}.
\end{align}
Similarly, the final form of common rate maximization problem with respect to the phase shift can be simplified to

\begin{align}
\max_{\varphi}&~{\widetilde{R}_{l,k,(m,n)}^c}=\nonumber\\
\max_{\varphi}&~\frac{2}{K}\text{Re}\{(\varphi_{l,k,(m,n)}^{(m,1)})^*q_{l,k,(m,n)}^{c,(m,1)}\}|\widetilde{\xi}_{l,k}^c|^2\widetilde{w}_{l}^{c,(m,m)},
\end{align}
where $q_{l,(m,n)}^{c,(m,1)}$ denotes the $m$-th element of vector $\bm{q}_{l,k,(m,n)}^{c}$, which is defined as $\bm{q}_{l,k,(m,n)}^{c}=\text{diag}(\mathbf{d}_{l,k,(m,n)})^H(\sqrt{1+\widetilde{\alpha}_{l,k}^c}\mathbf{w}_{l,k}^c(\widetilde{\xi}_{l,k}^c)^*+|\widetilde{\xi}_{l,k}^c|^2\mathbf{W}_{l}^c\mathbf{\widetilde{h}}_{l,k,(m,n)})$, and $\widetilde{w}_{l}^{c,(i,j)}$ is the $i$-th row and the $j$-th column element of $\mathbf{\widetilde{W}}_{l}^c$.

Therefore, the original optimization problem of maximizing the rate with respect to the PA position is transformed into an equivalent problem in terms of the phase shift for the $k$-th user on carrier $l$. This equivalent problem can be further reformulated as follows

\begin{align}
\max_{\varphi}~&\widetilde{R}_{l,k}^p+\tfrac{1}{K}\widetilde{R}_{l,k}^c=\label{phase_final}\\
\max_{\varphi}~&2\text{Re}\{(\varphi_{l,k,(m,n)}^{(m,1)})^*(q_{l,k,(m,n)}^{p,(m,1)}|\widetilde{\xi}_{l,k}^p|^2\widetilde{w}_{l}^{p,(m,m)}\nonumber\\
&+\frac{1}{K}q_{l,k,(m,n)}^{c,(m,1)}|\widetilde{\xi}_{l,k}^c|^2\widetilde{w}_{l}^{c,(m,m)})\}.\
\end{align}

To solve problem (\ref{phase_final}), it is necessary to align the phase shifts between the two product terms in curly braces. Consequently, the optimal phase is given by
\begin{align}
\angle \varphi_{l,k,(m,n)}^{(m,1),op}=&\angle (q_{l,k,(m,n)}^{p,(m,1)}|\widetilde{\xi}_{l,k}^p|^2\widetilde{w}_{l}^{p,(m,m)}\nonumber\\
&+\frac{1}{K}q_{l,k,(m,n)}^{c,(m,1)}|\widetilde{\xi}_{l,k}^c|^2\widetilde{w}_{l}^{c,(m,m)}).\label{op_s_s}
\end{align} 

Although the optimal phase shift has been determined, it cannot be directly used to derive the positions of the PAs. This is because the PA influences the phase shift solely through a change in its position, which modifies the propagation distance. This, in turn, uniquely affects each user’s propagation path, meaning the optimal phase shifts are fundamentally distinct and cannot be unified. Consequently, the PA position cannot be uniquely deduced from the optimal phase shift alone. To resolve this ambiguity, a one‑dimensional line search is employed to minimize the phase‑shift error. The problem can be formulated as follows
\begin{align}
	\min_{\chi_{(m,n)}^{PA}}\sum_{l=1}^F\sum_{k=1}^K|\angle \widetilde{\varphi}_{l,k,(m,n)}-\angle \varphi_{l,k,(m,n)}^{(m,1),op}|\label{phaseproblem},
\end{align} 
where $\angle \widetilde{\varphi}_{l,k,(m,n)}$ is the actual phase shift introduced to the $k$-th user on the $l$-th carrier by the $n$-th PA positioned at the $m$-th waveguide, and $\angle \varphi_{l,k,(m,n)}^{(m,1),op}$ is the corresponding ideal phase shift.

The phase shift error minimization problem (\ref{phaseproblem}) can be solved via a one-dimensional line search. Specifically, the optimization is performed sequentially. With the positions of all other PAs fixed, only a single PA is adjusted.  The optimal position of each PA is parameterized by a one‑dimensional variable defined over a search interval of length $s_{\Delta}$ with a step size $\delta_{\Delta}$. For every candidate position, the phase‑shift error objective (\ref{phaseproblem}) is evaluated. After completing the search, the position that minimizes (\ref{phaseproblem}) is selected as the optimal. This optimized position is then fixed, and the corresponding channel information is updated. The procedure is repeated sequentially for each remaining PA. The complete steps are summarized in Algorithm 1.

 \begin{algorithm}[!tbp]
 	\caption{{PA positioning algorithm based on minimizing phase shift error}}
 	\begin{algorithmic}[1]
 		\STATE $\textbf{Input}$ $h_{l,k,(m,n)}$, $\chi_{k}$, $\chi_{m}^{FP}$
 		\STATE $\textbf{Output}$ The PA position $x^{PA}_{op}$
 		\STATE  solve problem (\ref{largesc}) to get the initial positions of the PAs
 		\STATE  update auxiliary variables $\alpha_{l,k}^p$, $\alpha_{l,k}^c$, $\xi_{l,k}^p$, $\xi_{l,k}^c$
 		\STATE the optimal phase shift for a single user on a single carrier can be obtained from (\ref{op_s_s})
 		\STATE \textbf{Loop:} for all PAs
 		\STATE \quad optimal $x_{(m,n)}^{PA}$ and fix the other PAs.
 		\STATE \quad solve problem (\ref{phaseproblem}) by one-dimensional line search
 		\STATE \quad update the channel $\mathbf{h}_{l,k}$ using the latest PA position
 		\STATE \textbf{End loop}
 	\end{algorithmic}
 \end{algorithm}

\subsection{Waveguide Beamforming}
Once the PA positions are determined, the sum‑rate maximization problem reduces to a beamforming design problem. Since the PA positioning stage was primarily aimed at establishing a favorable wireless environment, the QoS requirements were not explicitly incorporated. Therefore, in this beamforming design phase, QoS constraints are explicitly introduced. Specifically, the Lagrange dual method is employed to optimize the beamforming design for multiple carriers. The Lagrange function of the sum-rate maximization problem can be expressed as:
\begin{align}
&\mathcal{L}(\mathbf{w}_{l,k}^p,\mathbf{w}_{l}^c,\mu,\lambda_k,\eta_{l},\zeta_{l})\nonumber\\
=&\sum_{l=1}^L\sum_{k=1}^K\widetilde{R}_{l}^p+\sum_{l=1}^LR_{l}^c\nonumber\\
&-\mu(\sum_{l=1}^L(||\mathbf{w}_{l}^c||^2+\sum_{k=1}^K||\mathbf{w}_{l,k}^p||^2)-P_{max})\nonumber\\
&+\sum_{k=1}^K\sum_{l=1}^L\lambda_{l,k}(\bar{R}_{l,k}^c-R_{l}^c)\nonumber\\
&+\sum_{k=1}^K\eta_{k}(\sum_{l=1}^L(r_{l,k}^c+\widetilde{R}_{l,k}^p)-R_{min})\nonumber\\
&+\sum_{l=1}^L\zeta_{l}(R_{l}^c-\sum_{k=1}^Kr_{l,k}^c)+\sum_{l=1}^L\sum_{k=1}^K\omega_{l,k}r_{l,k}^c,\
\end{align}
where $\mu$ is the Lagrange multiplier associated with the transmit power budget, $\lambda_{l,k}$ corresponds to the successful SIC constraint, $\eta_{k}$ is the multiplier for the transmission rate requirement, $\zeta_{l}$ is the multiplier for the common rate allocation, and $\omega_{l,k}$ ensures that the common rate allocated to each user is non‑negative.

Taking the derivative of the Lagrange function $\mathcal{L}$ with respect to $\mathbf{w}_l^c$ and setting it to zero yields:
\begin{align}
\tfrac{\partial \mathcal{L}}{\partial \mathbf{w}_l^c}=&2\sum_{k=1}^K\lambda_{l,k}\sqrt{1+\widetilde{\alpha}_{l,k}^c}\widetilde{\xi}_{l,k}^c\mathbf{h}_{l,k}\nonumber\\
&-2\sum_{k=1}^K\lambda_{l,k}|\widetilde{\xi}^c_{l,k}|^2\mathbf{h}_{l,k}\mathbf{h}_{l,k}^H\mathbf{w}_{l}^c\nonumber\\
&-2\mu\mathbf{w}_{l}^c=0.
\end{align}
Then the optimal common beamforming vector can be obtained as 
\begin{align}
\widetilde{\mathbf{w}}_{l}^c=(\sum_{k=1}^K\beta_{l,k}^c\mathbf{h}_{l,k}\mathbf{h}_{l,k}^H+\mu\mathbf{I})^{-1}(\sum_{k=1}^K\rho_{l,k}^{c}\mathbf{h}_{l,k})\label{wcclosed},
\end{align}
where $\beta_{l,k}^c=\lambda_{l,k}|\widetilde{\xi}_{l,k}^c|^2$ and $\rho_{l,k}^c=\lambda_{l,k}\sqrt{1+\widetilde{\alpha}_{l,k}^c}\widetilde{\xi}_{l,k}^c$.

Taking the derivative of the Lagrange function $\mathcal{L}$ with respect to $\mathbf{w}_{l,k}^p$ and setting it to zero yields
\begin{align}
	\tfrac{\partial L}{\partial \mathbf{w}_{l,k}^p}=&2(1+\eta_{k})\sqrt{1+\widetilde{\alpha}_{l,k}^p}\widetilde{\xi}_{l,k}^p\mathbf{h}_{l,k}\nonumber\\
	&-2\sum_{j=1}^K\lambda_{l,k}|\widetilde{\xi}^c_{l,j}|^2\mathbf{h}_{l,j}\mathbf{h}_{l,j}^H\mathbf{w}_{l,k}^p\nonumber\\
	&-2\sum_{j=1}^K(1+\eta_{k})|\widetilde{\xi}_{l,j}^p|^2\mathbf{h}_{l,j}\mathbf{h}_{l,j}^H\mathbf{w}_{l,k}^p\nonumber\\
	&-2\mu\mathbf{w}_{l,k}^p=0.
\end{align}
Then the optimal private beamforming vector can be obtained as
\begin{align}
\widetilde{\mathbf{w}}_{l,k}^p=(1+\eta_{k})\sqrt{1+\widetilde{\alpha}_{l,k}^p}\widetilde{\xi}_{l,k}^p(\sum_{j=1}^K\beta_{l,j}^{p}\mathbf{h}_{l,j}\mathbf{h}_{l,j}^{H}+\mu\mathbf{I})^{-1}\mathbf{h}_{l,k}\label{wpclosed},\
\end{align}
where $\beta_{l,j}^p=\lambda_{l,j}|\widetilde{\xi}_{l,j}^c|^2+(1+\eta_{j})|\widetilde{\xi}_{l,j}^{p}|^2$.

The Lagrange function involves multiple variables. For a more concise expression, like terms are combined, resulting in the following simplified form:
\begin{align}
	&\mathcal{L}(\mathbf{w}_{l,k}^p,\mathbf{w}_{l}^c,\mu,\lambda_k,\eta_{l},\zeta_{l})\nonumber\\
	&=\sum_{l=1}^{L} R_l^c \left( 1 - \sum_{k=1}^{K} \lambda_{l,k} + \zeta_l \right)\nonumber\nonumber\\
	&~+\sum_{l=1}^{L} \sum_{k=1}^{K}r_{l,k}(\eta_k-\zeta_l+\omega_{l,k}) \nonumber\\
	&~+\sum_{l=1}^{L} \sum_{k=1}^{K} (1 + \eta_k)  \widetilde{R}_{l,k}^p+\sum_{l=1}^{L} \sum_{k=1}^{K} \lambda_{l,k} \bar{R}_{l,k}^c\nonumber \\
	&~-\mu \sum_{l=1}^{L} \left( \| \mathbf{w}_l^c \|^2 + \sum_{k=1}^{K} \| \mathbf{w}_{l,k}^p \|^2 \right)
	+ \mu P_{max}
	- \sum_{k=1}^{K} \eta_k R_{\min}.
\end{align}

Building upon the closed-form expressions for the common and private beamforming vectors, it is observed that the Lagrange multipliers $\lambda_{l,k}$, $\eta_{k}$, and $\zeta_{l}$ exert a non‑monotonic influence on the sum‑rate objective; consequently, the Lagrange multipliers are updated according to the following rule:
\begin{align}
&\lambda_k^{(t+1)}=\max(0,\lambda_k^{(t)}+\delta_1(R_{l}^c-\bar{R}_{l,k}^c)),\\
&\eta_k^{(t+1)}=\max(0,\eta_k^{(t)}+\delta_2(R_{min}-\sum_{l=1}^L(r_{l,k}^c+\widetilde{R}_{k}^p))),\\
&\zeta_l^{(t+1)}=\max(0,\zeta_l^{(t)}+\delta_3(\sum_{k=1}^Kr_{l,k}^c-R_{l}^c)),\
\end{align}
where $\delta_1$, $\delta_2$, and $\delta_3$ are the positive step sizes.

Moreover, given that $\tfrac{\partial \mathcal{L}}{\partial r_{l,k}}=\eta_{k}-\zeta_{l}+\omega_{l,k}$, the complementary slackness condition of the Karush–Kuhn–Tucker (KKT) system gives  $\omega_{l,k}r_{l,k}=0$, while the KKT primal‑feasibility condition requires $\eta_{k}-\zeta_{l}+\omega_{l,k}=0$, yielding $\omega_{l,k}=\zeta_{l}-\eta_{k}$ together with  $r_{l,k}\geq0$. Since the dual feasibility demands $\omega_{l.k}$ is non-negative,  it follows that $\zeta_{l}-\eta_{k}\geq 0$. Furthermore, if $r_{l,k}\geq 0$ then $\zeta_{l}-\eta_{k}=0$ must hold. However, $\eta_{k}$ and $\zeta_{l}$ are continuous variables, which makes exact equality between them practically unattainable. To satisfy the condition $\zeta_{l}-\eta_{k}=0$ in the KKT sense, and considering that $r_{l,k}$ does not directly influence the objective function value, but merely serves as an auxiliary variable for common‑rate allocation to meet user rate requirements, a ratio‑based allocation scheme is introduced. The optimal auxiliary variable is then obtained as $r_{l,k}^{op}=\tfrac{(\eta_k-\zeta_l)^+}{\sum_{j=1}^K(\eta_j-\zeta_l)^+}R_{l}^c$ with $(a)^+$ is $\max\{0,a\}$.

Unlike the other multipliers, the condition 
$\tfrac{\partial \mathcal{L}}{\partial \mu}=\sum_{l=1}^L(||\mathbf{w}_{l}^c||^2+\sum_{k=1}^K||\mathbf{w}_{l,k}^p||^2)-P_{max}=0$ ensures feasibility. According to the Lagrange dual formulation, minimizing the dual function with respect to $\mu$ is equivalent to maximizing the original objective. Therefore, among all feasible values satisfying the derivative condition, the smallest possible 
$\mu$ is selected, i.e., 
\begin{align}
&\mu^{op}=\min\{\mu\geq\ 0:\sum_{l=1}^L(||\mathbf{w}_{l}^c||^2+\sum_{k=1}^K||\mathbf{w}_{l,k}^p||^2)=P_{max}\}.\
\end{align}
The optimal Lagrange multiplier $\mu^{op}$ can be determined via a bisection method.
At this stage, the beamforming optimization is completed. The complete beamforming optimization steps are summarized in Algorithm 2.

\begin{algorithm}[!tbp]
	\caption{{Beamforming Design Algorithm}}
	\begin{algorithmic}[1]
		\STATE $\textbf{Input}$ $h_{l,k,(m,n)}$, $\chi_{k}$, $\chi_{m}^{FP}$, $P_{max}$, $\mu_{max}$, $\mu_{min}$ 
		\STATE $\textbf{Output}$ The beamforming $\mathbf{w}_{l,k}^p$ and $\mathbf{w}_{l}^c$
		\STATE \quad update auxiliary variables $\alpha_{l,k}^p$, $\alpha_{l,k}^c$, $\xi_{l,k}^p$, $\xi_{l,k}^c$
		\STATE \quad update Lagrange multiple   $\lambda_{k}$, $\eta_{k}$,  $\zeta_{l}$, $r_{l,k}^{op}$
		\STATE $\textbf{Repeat:}$
		\STATE \quad update the beamforming according to (\ref{wcclosed}) and (\ref{wpclosed})
		\STATE \quad calculate the transmit power $p$
		\STATE \quad \textbf{if} $p > P_{max}$  
		\STATE \quad \quad $\mu_{max}=\tfrac{\mu_{max}+\mu_{min}}{2}$
		\STATE \quad \textbf{else} $p\leq P_{max}$  
		\STATE \quad \quad $\mu_{min}=\tfrac{\mu_{max}+\mu_{min}}{2}$
		\STATE \quad \textbf{end if}   
		\STATE $\textbf{Until}$ convergence is achieved or a predefined maximum number of cycles is completed.
	\end{algorithmic}
\end{algorithm}

\begin{algorithm}[!tbp]
	\caption{{Overall Sum-Rate Maximization for PA-Assisted RSMA}}
	\begin{algorithmic}[1]
		\STATE $\textbf{Input}$ $h_{l,k,(m,n)}$, $\chi_{k}$, $\chi_{m}^{FP}$, $P_{max}$
		\STATE $\textbf{Output}$ The sum rate $R_{sum}$
		\STATE \quad obtain initial PA positions  according to Algorithm 1.
		\STATE $\textbf{Repeat:}$ 
		\STATE \quad the positions of PAs are finely adjusted according to Algorithm 1.
		\STATE \quad optimize beamforming vectors according to Algorithm 2
		\STATE $\textbf{Until}$ convergence is achieved or a predefined maximum number of cycles is completed.
	\end{algorithmic}
\end{algorithm}

\begin{figure}[!t]
	\centering
	\includegraphics[width=2.5in]{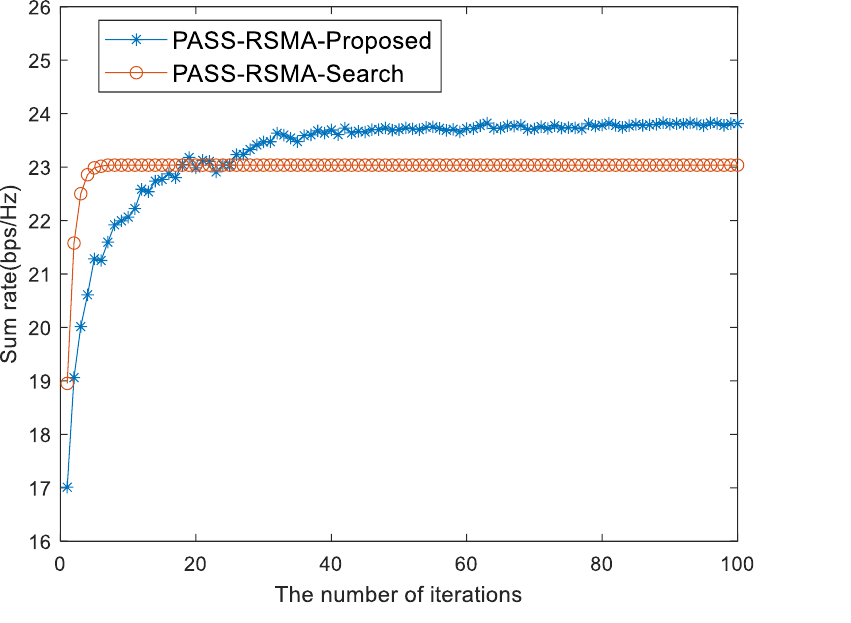}
	\caption{The sum rate versus the number of iterations with transmit power budget $P_{max}=25$ dBm and the deployment region size $D=6$ m.}\label{WANG_fig3}
\end{figure}

\subsection{Convergence and Complexity Analysis}
Although the proposed phase-shift-error-minimization method does not guarantee an increase in the objective function at every iteration, it ensures an overall upward trend and eventual convergence, as shown in Fig. 2. The complexity of the PA positioning scheme is $\mathcal{O}(MNKLI_{\text{iter}})$, where $I_{\text{iter}}$ denotes the number of line search iterations. In comparison, one-dimensional exhaustive search for sum-rate maximization shares the same order of complexity, i.e., $\mathcal{O}(MNKLI_{\text{iter}})$. Although the proposed scheme may require a larger number of iterations, each iteration involves only low-cost addition and subtraction operations, avoiding computationally expensive logarithmic evaluations. As a result, the overall computation remains efficient, enabling the algorithm to achieve near-optimal performance within practical time frames. Besides, the beamforming design entails a complexity of $\mathcal{O}(M^3)$, primarily due to the matrix inversion required for beamforming calculation. The complete steps are summarized in Algorithm 3.

\section{Simulation Results}
\begin{table}[!tbp]
\caption{Simulation Parameters}\label{simulation}\vspace{-0.3cm}
\centering
\scalebox{1}{
\begin{tabular}{c|c}
\hline \hline
Parameters & Values \\
\hline
The height of waveguides $Z^{PA}$ and $Z^{FP}$ & 3m \\
The absorption coefficient $\alpha$ & 0.01\\
The number of carriers $L$ & 3\\
The number of users $K$ & 4\\
The number of waveguide $M$ & 2\\
The carrier frequency $f_c$ & 28Ghz\\
The length of search step $\delta_{\Delta}$ & $0.001 \sim 0.005$m\\
The number of pinching antennas $N$ & $2 \sim 5$ \\
The deployment region size $D$ & 6-12m\\
The location of FP $(X^{FP}_{m},Y^{FP}_{m})$ & (0,-5,3), (0,5,3)m\\
The deployment region of users $(x_{k},y_{k})$ & $(0,-10) \sim (20,10)$m\\
The minimum rate requirement of user $R_{min}$ & 0.1 bps/Hz\\
The transmission power budget $P_{max}$ & $10-25$ dBm\\
Gaussian noise power $\sigma^2$ & -70 dBm \\
\hline
\end{tabular}}
\vspace{-0.3cm}
\end{table}
In this section, the proposed scheme (\textbf{PASS-RSMA-Proposed}) is evaluated. In this scheme, the sum rate is indirectly improved by minimizing phase shift errors through PA positioning. For comparison, it is assessed alongside several benchmark schemes designed to isolate the contributions of different factors. These include \textbf{PASS-RSMA-Search}, which was proposed in \cite{sun2025multiuser} to optimize PASS-SDMA system, where the sum-rate was directly maximized using a one-dimensional line search. The objective of this comparative study is to evaluate and contrast the efficacy of direct versus indirect optimization strategies. \textbf{PASS‑SDMA‑Search} is also considered, in which the same line‑search optimization is applied to a conventional SDMA system. This setup facilitates a direct performance comparison between RSMA and SDMA under PASS framework. Additionally, \textbf{PASS‑NOMA‑Search} adopts the above  line‑search optimization to obtain PA positions, but within a NOMA transmission structure. This setup facilitates a direct performance comparison between RSMA and NOMA under PASS framework. In order to demonstrate the effectiveness of the  fine-grained adjustment, \textbf{PASS-RSMA-Coarse} serves as a baseline in which PA positions are determined based solely on path loss, thereby highlighting the advantage of the proposed fine-grained optimization approach. Furthermore, to illustrate the benefits of the PASS architecture, a conventional hybrid beamforming system termed as \textbf{Hybrid-RSMA} is introduced for comparison. In this setup, the BS is located at the center of the users' movement area, equipped with $M$  RF chains, each connected to a subarray of $N$ antennas. The other simulation parameters are given in Table I.

 Fig. 2 depicts the convergence behavior with transmit power budget $P_{max}=25$ dBm and the deployment region size $D=6$ m.  Although the one‑dimensional line‑search PA position optimization scheme, which uses the sum rate as its objective, guarantees monotonic improvement per iteration and thus promotes rapid convergence, it also raises the risk of converging to a local optimum. In contrast, although the proposed phase shift error minimization scheme does not guarantee monotonic improvement per iteration, the convergence trajectory remains effectively guided toward an optimal state. Precisely, because the proposed approach permits temporary performance degradation, the scheme can escape local optima and get to a superior optimum. Consequently, the final solution tends to fluctuate stably within a bounded region around that optimum.

As depicted in Fig. 3, the proposed scheme outperforms all other baselines in terms of the achievable sum-rate with the transmit power budget and deployment region size $D=6$ m. The one-dimensional line search scheme \textbf{PASS-RSMA-Search} guarantees non-decreasing objective function values in each iteration but is prone to converge to local optima. In contrast, the proposed phase shift error minimization scheme \textbf{PASS-RSMA-Proposed} does not enforce monotonic improvement at every step. Instead, by allowing temporary exploration that can escape local optima, it ensures growth over the entire iterative process, ultimately converging to a superior solution. This explains why \textbf{PASS-RSMA-Proposed} achieves better performance than \textbf{PASS-RSMA-Search}, as observed in the figure. For \textbf{PASS‑RSMA‑Coarse}, although the PA positions are determined based on large‑scale channel information, the effect of the phase shift on performance is not involved. Specifically, the phase shifts introduced by coarse PA adjustment can lead to constructive or destructive function to the composite signal, thereby affecting overall performance.  This demonstrates the importance of fine-grained optimization in PA-assisted systems. Simulation results verify the effectiveness of the proposed PA positioning scheme. For \textbf{Hybrid‑RSMA}, although PASS operating principle is similar to conventional hybrid beamforming, the main advantage of PASS lies in its ability to adjust PA positions, thereby shortening the transmission distance. This reduction in distance mitigates path loss and consequently enhances the received signal strength. Compared to  \textbf{PASS-SDMA-Search}, which struggles to achieve effective interference mitigation in overloaded scenarios, RSMA manages interference effectively by dynamically adjusting the common stream. For \textbf{PASS‑NOMA‑Search}, while interference can be mitigated through SIC, the implementation of SIC in multi‑user scenarios introduces significant complexity and sensitivity to decoding order. In contrast, RSMA fundamentally transforms the interference management strategy by splitting user messages into common and private parts. Moreover, \textbf{PASS‑NOMA‑Search} exhibits lower performance than both \textbf{PASS‑RSMA‑Search} and \textbf{ PASS‑SDMA‑Search} due to its limited spatial multiplexing gain.

\begin{figure}[!t]
	\centering
	\includegraphics[width=2.5in]{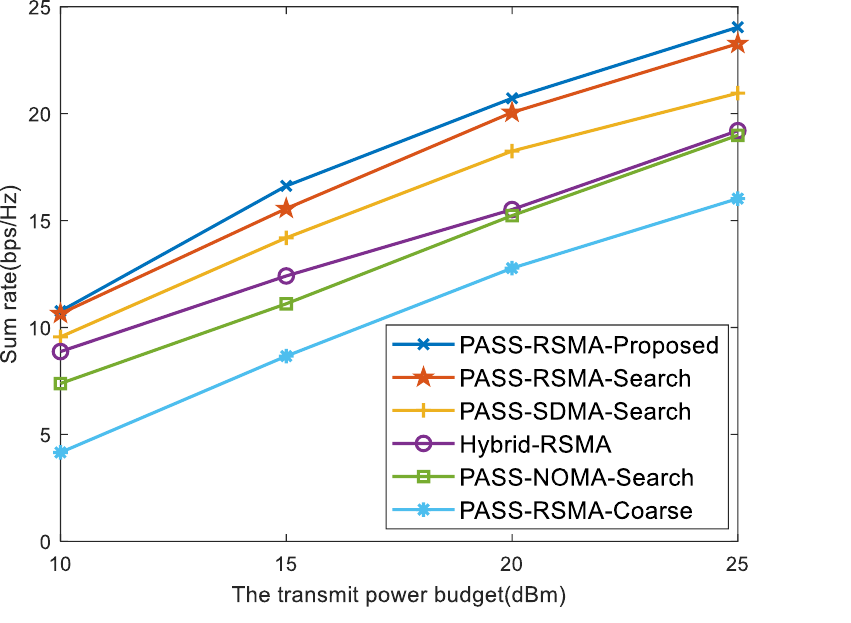}
	\caption{The sum rate versus the transmit power budget with $D=6$ m.}\label{WANG_fig2}
\end{figure}

In high-frequency communication scenarios, the short wavelength makes the channel phase highly sensitive to even small PA movement. Additionally, inaccuracies in the physical position of each PA on the waveguide introduce further uncertainties into the constructed channel. Therefore, robustness to imperfect CSI is crucial for every scheme. Fig. 4 illustrates the impact of imperfect CSI on the performance of the several schemes with the deployment region size of $D=$ 6 m. As illustrated in the figure, the RSMA demonstrates  the best performance against  imperfect CSI. This is because RSMA enjoys flexible interference management capability through splitting the data stream into common and private parts. In overloaded scenarios, SDMA cannot construct perfectly orthogonal beamforming, and imperfect CSI further exacerbates the difficulty of interference management. In NOMA, performance depends critically on the decoding order required for SIC, making it sensitive to imperfect CSI. Consequently, the performance of SDMA and NOMA schemes degrades more noticeably than that of RSMA, highlighting RSMA’s superior robustness and its stronger capability to compensate for impairments introduced by the PA configuration.

\begin{figure}[!t]
	\centering
	\includegraphics[width=2.5in]{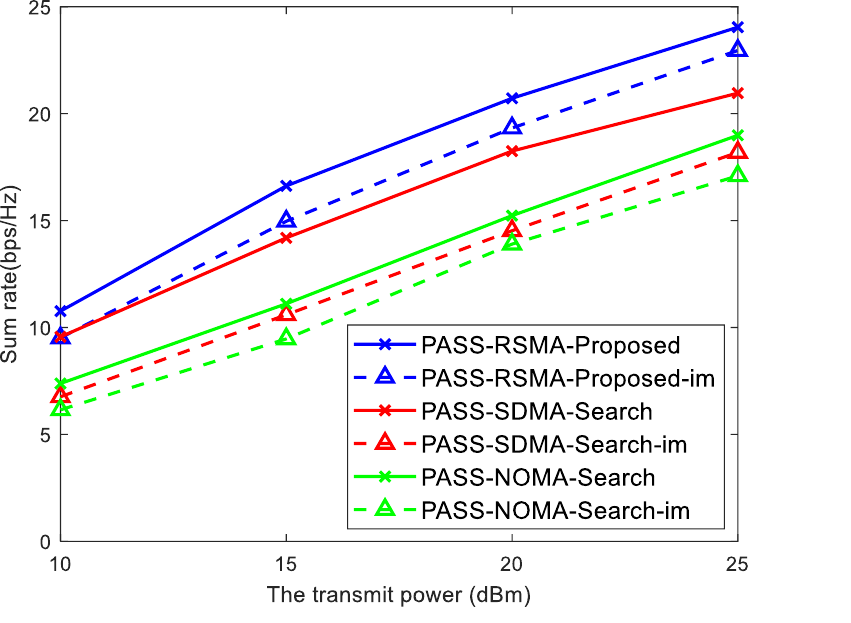}
	\caption{The sum rate versus transmit power budget under imperfect CSI with the deployment region size
		$D=$6 m.}\label{WANG_fig4}
\end{figure}

Fig. 5 illustrates the relationship between the number of PAs and the achievable sum rate with the transmit power budget $P_{max}=$25 dBm and deployment region size $D=$6 m. As shown, while the performance of all other schemes improves steadily as the number of PAs increases, \textbf{PASS‑RSMA‑Coarse} constitutes an exception. This is because a larger number of PAs provides greater freedom in shaping the channel, which in turn enhances the channel quality to improve the performance. However, the performance gain attainable through channel construction is inherently limited, leading to the observed steady‑state behavior. For  \textbf{PASS‑RSMA‑Coarse}, since it relies solely on distance to optimize PA positions, it can only mitigate path loss without optimizing the phase shifts of the PAs. Furthermore, when the number of PAs increases in \textbf{PASS‑RSMA‑Coarse}, these PAs do not participate in active position adjustment. Unlike fine-grained PAs, where adding more antennas typically enhances performance, simply increasing the number of PAs in \textbf{PASS‑RSMA‑Coarse} yields almost no user‑rate improvement. Simulation results demonstrate the necessity of fine-grained PA positions to construct channel for further optimization.

\begin{figure}[!t]
	\centering
	\includegraphics[width=2.5in]{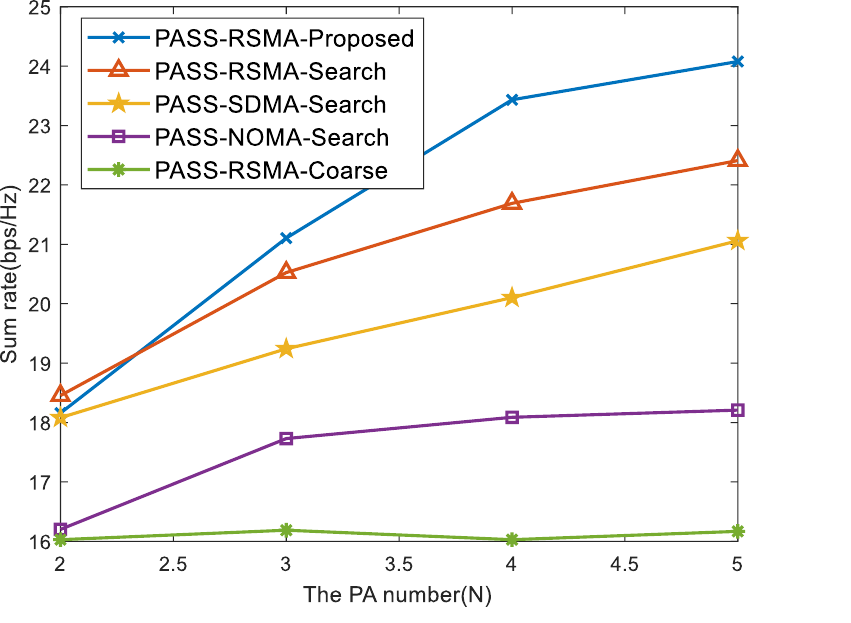}
	\caption{The sum rate versus the number of PAs on each waveguide with transmit power budget $P_{max}=$25 dBm and deployment region size $D=$6 m.}\label{WANG_fig5}
\end{figure}

As depicted in Fig. 6, which plots the sum rate against deployment region size with the transmit power budget $P_{max}=$25 dBm, the sum rate decreases as the deployment region size increases. Moreover, the performance gap between hybrid beamforming and PASS widens with the deployment region size increasing. This is because for conventional hybrid beamforming, the antenna positions are fixed and lack the adaptability inherent to PASS, resulting in a more significant performance degradation with deployment region size. In contrast, as the deployment region size increases, PASS can reposition its PAs to more favorable locations, thereby counteracting the increased path loss.

\begin{figure}[!t]
	\centering
	\includegraphics[width=2.5in]{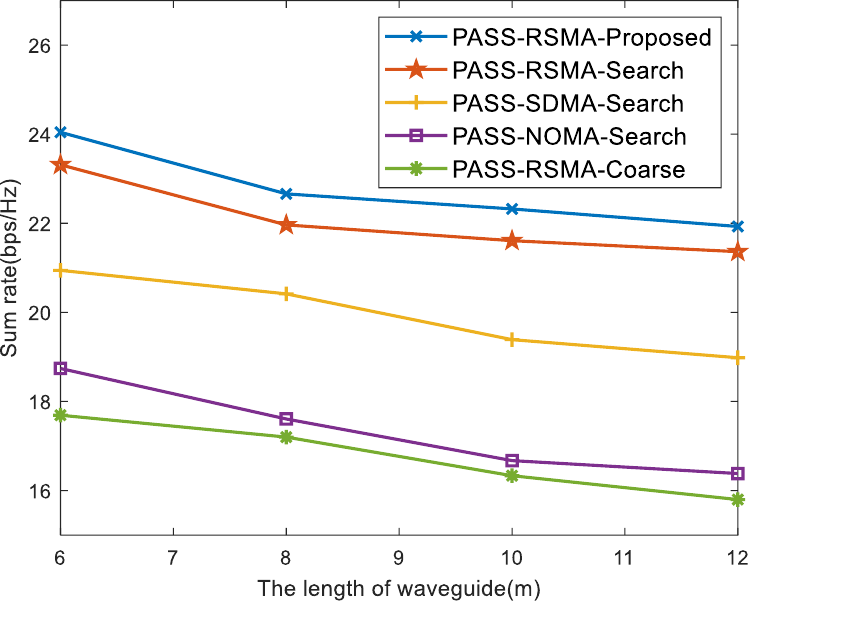}
	\caption{The sum rate versus the deployment region size with the transmit power budget $P_{max}=$25 dBm.}\label{WANG_fig5}
\end{figure}

As shown in Fig. 7, the sum rate is plotted against the search step length under different frequencies with the transmit power budget $P_{max}=$25 dBm and deployment region size $D=$6 m. The system achieves better performance at lower frequencies due to lower propagation loss. Furthermore, although decreasing the search step length generally improves performance, a smaller step size leads to higher computational complexity and longer execution time, necessitating a practical trade‑off. Moreover, performance does not improve indefinitely as the step size decreases; instead, the performance gain eventually saturates. This occurs because when the step size becomes sufficiently small, the PA can already find a position that yields near‑optimal performance, and further reducing the step length no longer significantly alters the PA placement. Therefore, selecting an appropriate step length is critical. Furthermore, the performance at high frequencies is more sensitive to the choice of search step length. This is due to the fact that, at higher frequencies, even a small PA displacement can induce a substantial phase shift, often spanning multiple signal cycles over the same movement distance. In contrast, at lower frequencies, the same physical displacement results in a much smaller phase variation.  Hence, the fine‑grained PA adjustment becomes increasingly essential in high‑frequency scenarios because higher carrier frequencies render the signal phase more sensitive to small PA displacements.

\begin{figure}[!t]
	\centering
	\includegraphics[width=2.5in]{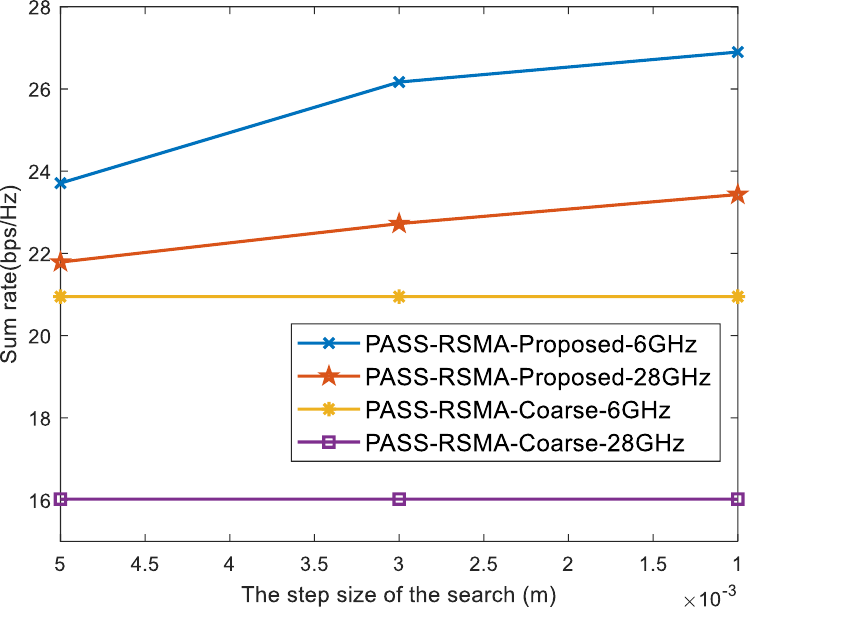}
	\caption{The sum rate versus the length of search step with the transmit power budget $P_{max}=$ 25 dBm and deployment region size $D=6$ m}\label{WANG_fig5}
\end{figure}

\section{Conclusion}
This paper investigates a multi-carrier multi-waveguide PA system under RSMA framework, with the objective of maximizing the sum rate in high-frequency scenarios. To address the performance degradation caused by the high sensitivity of high frequency signals to PA movement, we propose a phase shift error minimization method, in which a closed-form solution to the single carrier  single user case is first derived, and the PA position is then determined via a one-dimensional search based on minimizing phase shift error. Besides, a largrange dual founcation based scheme is employed to optimize waveguide beamfroming. Simulation results demonstrate that the proposed PA position optimization scheme is effective, delivering superior performance with low computational complexity. Moreover, the results confirm that fine PA position adjustment is essential, and its importance becomes even more pronounced in higher‑frequency scenarios. Furthermore, in PA-assisted systems, RSMA demonstrates significantly stronger robustness to imperfect CSI than SDMA and NOMA schemes.
\vspace{-0.2cm}

\bibliographystyle{IEEEtran}
\bibliography{reference}

\end{document}